# Toward Trustworthy Urban IT Systems: The Bright and Dark Sides of Smart City Development

Jungheum Park, Hyunji Chung

*Abstract*— In smart cities built on information and communication technology, citizens and various IT systems interoperate in harmony. Cloud computing and Internet-of-Things technologies that have been developed for a long time are making modern cities smarter. Smart cities can have a positive impact on citizens, but they can also make cities dangerous. Today, with the emerging reality of smart cities, this paper looks at both the bright and dark sides and provides a foundation for supporting work-related tasks of IT professionals as well as non-IT experts involved in urban design and development.

## I. Introduction

The development of cloud computing technology has ushered in the era of the Internet of Things (IoT). The "Smart City," where everything in the city is connected to the Internet and works organically, will soon unfold in front of our eyes. Along with the Fourth Industrial Revolution, the likelihood of cities becoming smarter is expected to increase. Citizens, who are at the heart of smart cities, can use various services and facilities provided by the city. Through collaborative services, social networking services, delivery services, and home automation, citizens can communicate with one another and live a more convenient life than in the past. Moreover, urban infrastructure, such as energy, transportation, healthcare, and transportation systems, can be smarter through sensing and network technology. Becoming smart in cities can improve the quality of life for citizens and enable cities to automatically heal themselves.

However, everything has both bright and dark sides, like the two sides of a coin. The dark side includes the risk that smart cities will collect too much information about individuals, which can infringe on individual privacy and human rights. Digital devices with security vulnerabilities can also leave the entire city open to cyberattacks [1]. In addition, a balance between usability and security to make smart cities safe and digital dependence for the mental and psychological health of citizens must be considered. With the largest cities around the world building smart cities, we point out the positive and negative aspects of smart cities for people.

## II. Orchestrating Smart City Ecosystems

People and things in a smart city generate large amounts of data per day, and the data are stored in or transferred through the cloud, as shown in **Fig. 1**. This section looks at each type of component and device to determine if they function organically to make the smart city work. This paper examines how life has become smarter from the perspective of citizens and how facilities and systems have changed people's lives from the perspective of cities. We emphasize matters that directly affect people's lives, rather than considering deep technical levels.

### A. People's viewpoint: smarter and customized services

In this subsection, we examine how cloud-based services have changed the lives of individuals.

*1) Collaboration:* People need to collaborate with others in their lives. As the cloud computing environment matures, workers can collaborate with others anywhere, anytime, and see each other in real time. This has the advantage of promoting efficiency. In particular, the use of online collaborative services has rapidly increased during the recent COVID-19 pandemic.

*2) Social networking:* In 2020, 3.81 billion users use SNS such as Facebook and WhatsApp. People can post text and upload photos/videos via SNSs. The contents are stored on service providers' servers and shared with others in real time. SNSs, based on cloud technology, enable people to make friends across borders and communicate with one another online.

*3) Photo-syncing:* People no longer back up photos taken with a smartphone on their personal computer. Instead, they utilize cloud-based photo storage services, such as Google Photo. Services are in the process of evolving to perform object recognition, automatic classification, and topic identification beyond storing pictures [2].

*4) Ride-sharing:* Services that allow multiple people to share automobiles, bicycles, and electric bikes have appeared in recent years. Multiple users can borrow and return vehicles from different locations simultaneously. It is also possible to check where they are available in real time. Sensors attached to each device collect various information, such as the current location, and then transmit it through wireless networks to the cloud in real time.

*5) Delivery:* With the increasing number of single-person households and the occurrence of a pandemic, people have been increasingly ordering food or goods through delivery services. Systems that deliver human-ordered food to restaurants and manage order history also operate based on cloud computing. In the past, people had to go directly to the market to buy ingredients, but food delivery services help people live more comfortably.

*6) Home automation:* In the past, people had to manage basic tasks, such as washing, eating, and resting at home, themselves. In a cloud-based smart home environment, coffee is made by voice command, refrigerators automatically order groceries, and kids enjoy entertainment through IoT devices such as virtual reality (VR) headsets. People can weigh themselves and send the information to the cloud to automatically analyze how much more they need to exercise and whether there is a change in muscle mass. Cloud computing makes the home environment smarter and more comfortable.

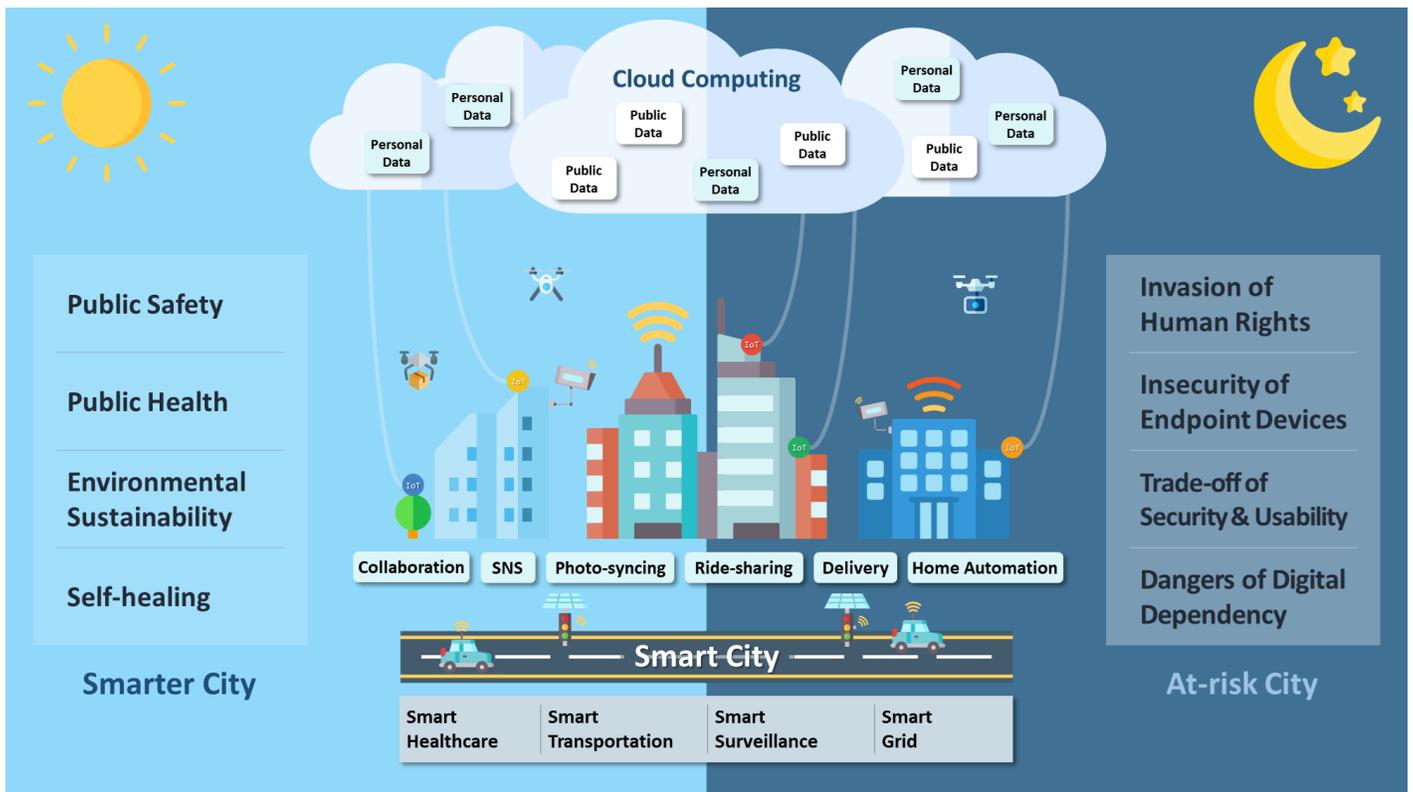

*Figure 1. Smart city ecosystems based on IoTs and clouds: the bright and dark sides of smart city development*

*B. City's viewpoint: smarter urban facilities and systems*

This subsection examines how smarter systems have changed traditional concepts in society.

*1) Smart healthcare:* A cloud-based smarter city can protect people's health and save lives. People may use a fitness tracker while jogging that transmits the measured movement route, steps, heart rate, and stress index to the cloud to manage their health through data analysis. In the past, health centers and hospitals were used for this purpose, but people can now manage their own health without seeing medical staff. In addition, sensing and cloud computing technologies are combined to transmit the status of critically ill patients equipped with medical equipment to medical staff in real time, thereby saving patients' lives [3].

*2) Smart transportation:* Transportation systems supported by IoTs and clouds can improve efficiency and safety in public live. Hangzhou, China, one of the top three smart cities in China, is a prime example. Alibaba has been developing a platform called "City Brain" since 2015, and it has been applied directly in Hangzhou [4]. In 2016, traffic jams were the biggest problem in Hangzhou, as the fifth largest city in China. In response, Alibaba mounted IoT devices on 104 traffic lights and collected data in the cloud. As a result, the traffic jam time was reduced by 15%, and the city brain platform was expanded with 1,300 traffic lights and 3,700 traffic cameras.

*3) Smart surveillance:* As in the case of several countries, intelligent CCTV technology such as the Domain Awareness System (DAS), NGI (Next Generation Identification, FBI), and Tian Wang of China, are widely used as next-generation social safety systems for finding suspects in public places, detecting and tracking them, and responding to dangerous situations. A case in point is China's recent arrest of five suspected criminals in two months at the concert hall based on face recognition technology. All CCTVs, various security sensors, and big data information are organically analyzed, linked, and developed into an integrated framework that responds to various social safety threats. In other words, instead of intelligent CCTVs with a fragmentary response and follow-up to sporadic social safety risks, they are expanding into intelligent urban security platforms that systematically protect citizens' lives and property by tying the entire city in a single domain [5].

*4) Smart grid:* Through smart grids, energy efficiency can be maximized by technology that can determine electricity usage, supply amount, and the state of power lines, producing as much electricity as necessary, and reducing electricity that is stored and re-supplied through accumulators when necessary. Smart cities, which combine smart grid technology, have attracted the attention of many countries and companies. For example, South Korea prepared a national roadmap for smart grid infrastructure in January 2010. Under the roadmap, the government aims to invest USD 26.8 billion by 2030 to establish a nationwide network of smart grids [6]. The focus in the United States has been on ways to commercialize smart grids through business models, secure stable supplies of energy, introduce renewable energy smoothly, and use it effectively. In addition, smart grid policies are being implemented in various countries, such as Germany, France, China, and the United Arab Emirates. As a result, smart city projects are being carried out in various cities, including Masdar in the United Arab Emirates and Maui Island in Hawaii.

## III. THE BRIGHT SIDE

An intelligent city can detect, resolve, and prevent crimes or natural disasters that threaten the safety of the city and its residents.

### A. Public safety

Building infrastructure that can maintain safety and prevent accidents is an important factor in smart city development.

In the case of Richmond, Virginia, the crime rate has recently dropped to its lowest level in 26 years owing to the convergence of business information, predictive analysis, data mining, and geographic information system (GIS) data on the basis of crime probabilities that occur in a certain space and time [7]. The data collected from the police records management system can be used to discover crime-causing hotspots by combining various factors such as time, weather, and public events, to prevent criminal incidents.

The City of London, England, uses an integrated system consisting of a CCTV to set up on the road and detect and track criminal activity [7]. Such CCTVs include automatic recognition of license plates as well as tracking of the entries and exits of cities.

China is at a level at which it can predict the number of people in a particular zone, their names and patterns of behavior after a certain period of time. More than 40,000 face recognition CCTVs were installed in 14,000 villages in Sichuan Province. In addition, the government has established a public surveillance network, in which Chinese public security and residents work together to monitor situations at sites in real time by connecting the residents' TVs, mobile phones, and public security systems. Since its implementation, the number of crimes committed in Sichuan Province has decreased by 50%, and the rate of arrests has increased by 50%, proving its effectiveness. Therefore, all CCTVs in China are being integrated into a central data-sharing platform at the national level and expanded to areas where surveillance has been difficult in the past [8].

### B. Public health

With the recent outbreak of COVID-19, citizen data management systems have been attracting more attention as a part of smart city development [9]. In Korea, the "COVID-19 Epidemiological Survey Support System," which has been effectively used by quarantine authorities since March 2019 to track the movements of confirmed patients and identify their stay and travel history, brought the smart city technology platform to quarantine sites. In other words, the technology-supported quarantine system automated the epidemiological investigation procedures of the Korea Centers for Disease Control and Prevention by collecting and analyzing the location and activity data of residents in large urban areas. Thus, based on the quarantine system that played a role in curbing the rapid spread of the virus, Korea created a system that allows people who are infected to be quickly detected and treated.

### C. Environmental sustainability

Air pollution is an emerging problem in many cities. According to the World Health Organization (WHO), approximately 7 million people, or one eighth of the world's total death toll, died in 2012 owing to air pollution exposure. Sensors installed throughout the city detect local humidity, dust levels, hazardous chemicals, barometric pressure, and other factors in minutes and report them to the central system. Local governments can use this information to justify controlling traffic and industry in highly contaminated areas and establishing budgets for improving air quality.

In addition, overflowing garbage can be harmful to urban hygiene. However, sending someone to check whether the trash can is full can be a waste of time. With solutions for monitoring the status of trash cans, garbage collection personnel can travel shorter distances and use less fuel. It is also possible to reduce the number of garbage vehicles in operation. When a smart trash can is emptied, it generates a notification through a wireless network. This reduces unnecessary collection while keeping trash cans from overflowing [10].

### D. Self-healing

Beyond environmental sustainability, an intelligent city can facilitate self-healing processes from possible accidents or disasters occurring in urban facilities and systems. In areas with high levels of fine dust, a system to lower the concentration is operated for a quick response. Another system stores electrical energy in advance when electricity prices are low and then uses it when the demand for electricity is soaring. This can solve the power supply imbalance [11]. As an example of disaster recovery and management, in Japan, the smart Business Continuity Plan (BCM) project, which was applied in the Shinagawa/Tamachi area of Minato, was launched for the 2020 Tokyo Olympics. It has established disaster prevention functions that can withstand natural disasters such as earthquakes, and disaster recovery plans are also being pursued [12].

## IV. THE DARK SIDE

Unfortunately, it is not only bright in a smarter city. This section looks at the negative side that can emerge during the construction and operation of smart city infrastructure. Specifically, we would like to emphasize the cybersecurity and privacy perspectives that IT professionals should be more interested in when developing a trustworthy smart city.

### A. Invasion of privacy and human rights

Advanced urban infrastructures have the advantage of preventing crimes and helping to use energy and other resources efficiently. To build such infrastructure, it is necessary to collect data on individuals and organizations. Some raise concerns that private companies with technology could hand over the initiative to operate urban systems such as healthcare, public security, transportation, and environmental systems. Moreover, surveillance and privacy concerns create psychological stress, owing to the fear that smart cities will merge into the "monitoring of capitalism." Surveillance capitalism, a concept established by Shosana Zubov, an honorary professor of business administration at Harvard University, not only makes profits by utilizing human behavior data but can also use technology to guide people in a desired direction. When developing smart cities, issues such as how to share collected data, distribute surplus generated in the process, and balance the protection and utilization of sensitive personal information should be dealt with in depth, but some countries are complacent under the assumption that the Fourth Industrial Revolution is a trend [13]. A sophisticated surveillance system led by companies could be more dangerous than China, which is an outright "big brother" [14].

Privacy concerns can also arise during the current pandemic [9]. For example, quarantine systems are gaining the ability to further threaten privacy through various information collection methods, including self-quarantine safety protection apps for tracking infected people, QR code-based entry logging services, identity verification services based on face recognition and heat sensing techniques, credit card providers, public CCTVs, and public transportation services. The more non-face-to-face methods are strengthened in the era of infectious diseases, the more automated smart cities will be expanded through big data collection and analysis. In the process, it is worrisome in terms of privacy and human rights to enforce the reality of digital control, which records and analyzes almost all citizens' activities in terms of their health and safety. In particular, it would be even more threatening if specialized platforms of certain IT companies became fundamental parts of a smart city.

### B. (In)Security of endpoint devices

Recently, a variety of smart devices have become available. It is unclear whether endpoint devices will be released with proven security. In the recent case of the Mirai Distributed Denial of Service (DDoS) attack, the major factor that enabled the cyberattack was that device owners usually do not change the default passwords of the devices they buy [15]. This is an example of the exploitation of an unspecified number of IoT devices that did not comply with basic security settings. In future smart cities, there will be much more diverse endpoints available than today; therefore, the entire city is more likely to be exploited for cyberattacks, unless there is a process to verify the security of the devices [16].

### C. Trade-off of security and usability

In cloud computing environments, accessing cloud-based storage requires a process of verifying whether you are an authorized user. As the number of cloud services increases, the amount of account information that needs to be managed also increases. There is a limit to the information that people can remember, but the number of services that need to be used is increasing. These changes have made it necessary to consider security for user convenience. Recently, in addition to passwords or passcodes, authentication methods using biometric information, such as fingerprint, voice, and face, have been applied for better convenience. Future cloud-based smart cities will need access to more diverse services and systems, and the authentication methods required to access them will also vary depending on their operating environments. Therefore, it is necessary to design services that consider the convenience of citizens as well as cybersecurity to protect their information.

### D. Digital dependency

One of the most worrisome aspects of the smart city environment is that citizens and cities are too dependent on technology. Since the advent of IT and networks, people have become addicted to digital devices and online services in their daily lives. In the past, there were many things that people had to remember, but, in recent years, it has become natural to store information on digital devices instead. To solve this dependency problem, people need to change their behaviors to avoid relying too much on the digital world mentally and psychologically. As digital dependence grows, interconnections in a broader city perspective are exposed to internal and external threats through ever-increasing potential attacks. Even if it is not threatened by cyberattacks, the whole city and people depending on technology may be paralyzed, even if a small part of its complex systems malfunctions unexpectedly. Therefore, research on social and technological solutions is needed to avoid increasing digital dependence in smart cities too much [17].

## V. CONCLUSION

Smart cities are already unfolding before our eyes. As described in this paper, smart city development has many advantages, such as public safety, public health, environmental sustainability, and self-healing. However, it is necessary to recognize that there is a contrasting dark side of the coin. In particular, this paper describes the representative dark aspects that can directly affect people. By emphasizing the ambivalence of smart cities, it raises awareness among IT professionals and delivers understandable levels of basic knowledge to non-IT experts involved in urban design and development.

As a starting point for further research, this study can be extended to a two-sided analysis of technical and social perspectives on individual cloud/edge servers and endpoint devices that make up a smart city. Furthermore, research activities, including but not limited to the development of security policies/guidelines for endpoint devices, the anonymization of sensitive information, and the legislation of privacy, would be meaningful to eliminate or mitigate the negative factors presented in this paper.

If all stakeholders understand the dark aspects inherent in most technologies being operated around people and complement those aspects to reinforce their bright aspects, we believe that it is possible to build a trustworthy smart city.

JUNGHEUM PARK is a research professor with the School of Cybersecurity, Korea University, Seoul, South Korea. His research interests include security, privacy, and digital forensics. Contact him at jungheumpark@korea.ac.kr.

HYUNJI CHUNG is a research professor with the School of Cybersecurity, Korea University, Seoul, South Korea. Her research interests include IoT forensics and smart city. Contact her at localchung@gmail.com (corresponding author).